\documentclass[aps,prl,reprint,groupedaddress,noeprint]{revtex4-2}

\usepackage{graphicx}
\usepackage{dcolumn}
\usepackage{bm}
\usepackage{hyperref}

\begin{document}


\title{Detecting dark objects in the Solar System with Gravitational Wave observatories}
	
	\author{Valentin Thoss}
	\email{vthoss@mpe.mpg.de}
	\affiliation{Universitäts-Sternwarte\\
		Ludwig-Maximilians-Universität München\\
		Scheinerstr. 1\\
		81679 Munich, Germany}
	\affiliation{Max-Planck-Institut für Extraterrestrische Physik\\
		Gießenbachstraße 1\\
		85748 Garching, Germany}
	\affiliation{Excellence Cluster ORIGINS\\
		Boltzmannstraße 2\\
		85748 Garching, Germany}
	
	\author{Abraham Loeb}
	\affiliation{Astronomy Department\\ Harvard University\\60 Garden St.\\ Cambridge, MA 02138, USA}
    
\date{\today}

	\begin{abstract}
        Dark objects streaming into the solar system can be probed using gravitational wave (GW) experiments through the perturbations that they would induce on the detector test masses. In this work, we study the detectability of the resulting gravitational signal for a number of current and future GW observatories. Dark matter in the form of clumps or primordial black holes with masses in the range $10^7$-$10^{11}\,\rm{g}$ can be detected with the proposed DECIGO experiment.
	\end{abstract}


\maketitle

\section{Introduction} 
	
	Gravitational wave (GW) observatories such as LIGO-Virgo-KAGRA have opened a new window on the universe by their ability to detect minuscule deformations of spacetime caused by the inspiral and merger of black holes binaries and neutron stars \cite{FirstGW,FirstNS,GWTC3}. They do so by interfering light beams sent between the detector test masses.

    During their operation, GW experiments are subject to a number of different noise sources, some of which lead to movement of the test masses between which the deformations of spacetime are measured. Such perturbations, if they are unpredictable, generally pose a limit to the sensitivity of an experiment \cite{LIGONoise}.

    However, the movement of the detector test masses can also lead to a detectable signature. In this work we are interested in gravitational perturbations induced on the test masses by fast transients of objects passing through the solar system. These objects could be primordial black holes (PBHs), making up the entire dark matter \cite{Carr2021,Escriva}. However, the results from this work generally apply to any object flying through the solar system unbound, with velocities beyond the escape speed.

    Similar analyses have been performed in the past for the case of dark matter clumps \cite{Adams,Seto,Cerdonio,Hall,Baum}. However, these studies either focused on a single detector such as LISA, included forces beyond a pure gravitational interaction, or did not take into account the full noise profile of the experiments. Our contribution is twofold. On the one hand, we provide a general way to estimate the detectability of an object passing through the solar system with any velocity, mass or distance. On the other hand, focusing on the case of dark matter clumps, we provide a general overview on the detectability for a number of current and planned GW detectors. We will also pay special attention to PBHs with a mass $M<10^{15}\,\rm{g}$ which have previously been ignored due to their presumed rapid evaporation. However, recently this view has been challenged by the so-called 'memory burden' effect \cite{Dvali2018,Dvali2020,Thoss,Alexandre,Zell,Montefalcone,Dondarini} and it is therefore worthwhile to study the scenario of such light PBHs as well.

    \cite{Badurina} studied the detectability of dark matter with proposed atom gradiometers. Their detectable mass window ($M\in[10^6,10^{10}]\,\rm{g}$) is similar to what we find for the planned BBO/DECIGO experiments and their results are thus complimentary to our work. We also want to emphasize that there have been numerous other works on the role that dark matter clumps or primordial black holes could have in the solar system \cite{Seismic,Geller,Crater1,Crater2,GNSS,EarthMoon,ThossSolar,Seismic}.

    Our paper is structured as follows. We first describe how we compute the signal of the fly-by for any gravitational wave detector. In the following we investigate detection prospects in terms of the detection volume of each experiment as well as the interstellar mass density that they are sensitive to.
	
	\section{Computation of the signal} 
	
	An object with mass $M$ that traverses through the solar system with speed $v$ will exert a gravitational force on the test mass of a GW detector, leading to a time-dependent acceleration
    \begin{equation}
        a(t) = \frac{GMR}{\sqrt{R^2+v^2t^2}^3}\,,
    \end{equation}
    in the direction towards the point of closest encounter at a distance of $R$. The spectral density of such a signal can be computed analytically to be
    \begin{equation}
        a(\omega) = \frac{2GM\omega}{v^2}K_1(\omega R/v)\,,
    \end{equation}
    where $K_1$ denotes the first-order modified Bessel function of the second kind. Note that $K_1(x)\approx1/x$ for $x\ll1$ and thus $a(\omega)=\frac{2GM}{vR}$ for $\omega\ll v/R$. 
    
    For a detector with multiple test masses $i$, located at positions $\mathbf{r}_i$, the different masses will be perturbed at a slightly different time $t_i$ and distance of closest encounter $R_i$. The spectral density of the relative acceleration $\delta a_{ij}(\omega)$ between a pair of test masses $i$ and $j$ can be computed by projection on the vector between the two test masses $\mathbf{r}_i-\mathbf{r}_j$:
    \begin{equation}
        \delta a_{ij}(\omega) = \left(a_i(\omega)\mathbf{\hat{a}_i}-a_j(\omega)\mathbf{\hat{a}_j}e^{i\omega \Delta t}\right)\cdot\frac{(\mathbf{r}_i-\mathbf{r}_j)}{L}\,.
        \label{eq:acc_general}
    \end{equation}
    Here $\Delta t=t_i-t_j$ is the difference between the time of closest encounter for each detector test mass and $L=|\mathbf{r}_i-\mathbf{r}_j|$ is the arm length of the detector. By $\mathbf{\hat{a}}_{i}$, we denote the normalized vector pointing from the position of the test mass to the point of closest encounter.
    
\begin{figure}
		\centering
        \includegraphics[width=0.7\linewidth]{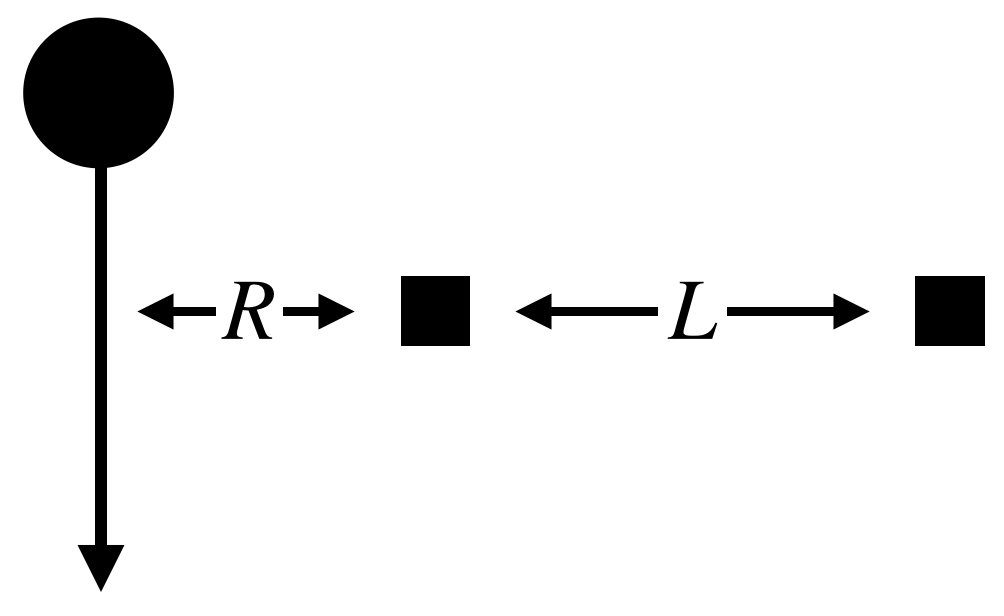}
		\caption{Visualization of a fly-by at an impact parameter $R$ near a GW detector of arm length $L$}
		\label{fig:sketch}
	\end{figure}

    To gain a better understanding of Equation~\ref{eq:acc_general}, we consider a single detector arm with two test masses. The largest perturbation will be achieved if the compact object moves perpendicular to the axis of the detector arm, as sketched in Figure~\ref{fig:sketch}. In this case $\Delta t=0$ and Equation~\ref{eq:acc_general} reduces to:
    \begin{equation}
        \delta a = \frac{2GM\omega}{v^2}\left(K_1(\omega R/v)-K_1(\omega (R+L)/v)\right)
        \label{eq:acc}
    \end{equation}
    If the compact object passes within close proximity of the test mass ($R\ll L$, near-field limit), then that mass will be perturbed much stronger than the other one and we get to good approximation
    \begin{equation}
        \delta a_{\rm near}=\frac{2GM\omega}{v^2}K_1(\omega R/v)\,\,
        \label{eq:acc_near}
    \end{equation}
    for the geometry depicted in Figure~\ref{fig:sketch}.
    If the distance between the compact object and the test masses is much larger than the detector size ($R\gg L$), we only get a tidal effect, where the perturbation is suppressed by the factor $L/R$:
    \begin{equation}
        \delta a_{\rm tid}\approx \frac{2GM\omega L}{R v^2}K_1(\omega R/v)
        \label{eq:acc_far}
    \end{equation}

    In general, any other geometry of the encounter will produce a smaller perturbation $\delta a$. However, a gravitational wave detector usually has more than two test masses. For an L-shaped detector like LIGO with 3 test masses, the measured quantity is actually $\delta a_{12}(\omega)-\delta a_{13}(\omega)$. Here, the index 1 indicates the corner test mass where the signals from the two arms interfere. Therefore, the measured perturbation can in principle be larger by up to a factor of two compared to Equation~\ref{eq:acc}. It is useful to evaluate the probability distribution $p(\delta a(\omega))$ due to the space of possible geometries of the encounter. Using Monte-Carlo simulations we sample this distribution for a LIGO-like 'L'-shaped detector, as well as for a triangular setup as is planned for LISA, BBO and DECIGO. We find that the following expression gives an excellent fit to the mean perturbation strength both for $R\ll L$ and $R\gg L$:
    \begin{equation}
        \langle\delta a(\omega)\rangle= \frac{2GM\omega L}{v^2(\alpha R+\beta L)}K_1(\omega R/v)
    \end{equation}
    For an L-shaped detector we obtain $\alpha=1.3$ and $\beta=1.76$ and for a triangular shape we get $\alpha=1.5$ and $\beta=2.0$. The maximum deviation of the fit from the simulated data is observed for $R=L/2$, when the expression above underestimates the mean perturbation strength by a factor of $\sim$ 1.6. In the following, we will make use of this expression to obtain estimates of the signal-to-noise ratio (SNR) for existing and planned GW observatories.
    
    To compute the spectral density of the gravitational wave strain $h(\omega)$, we recall that $h= 2\delta x/L$. From this and $\delta x(\omega)=\delta a(\omega)/\omega^2$ we get the strain
    \begin{equation}
        h(\omega)=\frac{4GM}{\omega v^2(\alpha R + \beta L)}K_1(\omega R/v)\,,
        \label{eq:meanstrain}
    \end{equation}
    which can be used to estimate the SNR for any GW experiment, as
    \begin{equation}
        \mathrm{SNR}=2\left(\int\mathrm{d}f\,\frac{h(f)^2}{P_n(f)}\right)^{1/2}\,,
        \label{eq:snr}
    \end{equation}
    where $P_n(\omega)$ denotes the power spectral density of the detector noise for the respective observatory. Note that for the case of gravitational waves, $P_n(f)$ is replaced by the spectral strain sensitivity $S_n(f)$. The latter usually accounts for the angular dependence of the response of the detector. In addition, it includes a frequency-dependent suppression for gravitational waves with wavelengths which are comparable or smaller than the arm length of the detector \cite{GWSC2}. In our case this suppression is not applicable and the angular dependence is accounted for in the signal by the factors $\alpha$ and $\beta$. Therefore, we can directly use the detector noise $P_n(f)$ to estimate the SNR. 
    
\begin{figure}
		\centering
        \includegraphics[width=\linewidth]{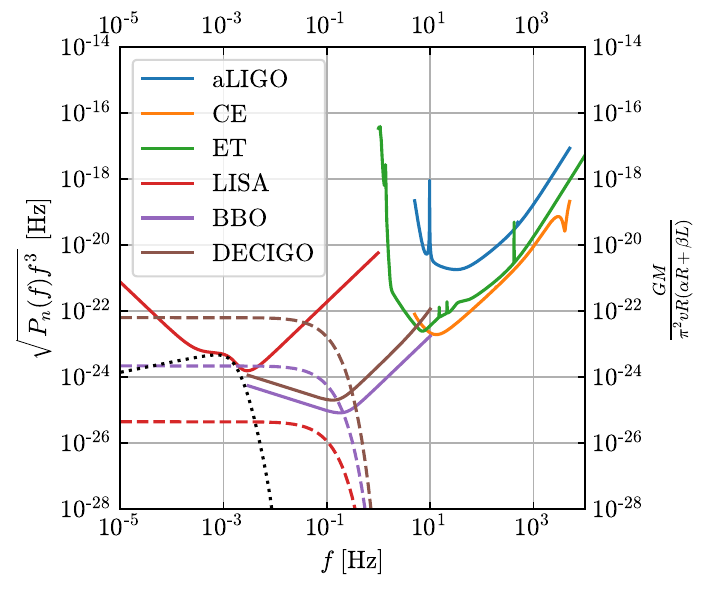}
		\caption{Strain sensitivity of various existing and proposed GW experiments. The y-axis has been rescaled such that the signal of an encounter with given parameters $M$, $R$ and $v$ would appear as a horizontal line for $f<v/R$, dropping quickly for $f>v/R$. The area between it and any sensitivity curve is proportional to the signal-to-noise ratio for that experiment. Note that in the limit of a close encounter, $R<L$, the signal amplitude is dependent on the detector arm length and must be computed for each experiment separately. The dashed lines shows the signal for an object with $M=10^{10}\,\rm{g}$, $v=300\,\rm{km\,s}^{-1}$ and $R=10^{3}\,\rm{km}$ for the detectors LISA, BBO and DECIGO in the same colors as the strain sensitivity. The dotted black line indicates the binary confusion noise, taken from \cite{LISAConfusion}.}
		\label{fig:straintidal}
	\end{figure}

    The data for our study is taken from \cite{Schmitz} for the detectors aLIGO (design sensitivity), Einstein Telescope (ET), LISA, Big Bang Observer (BBO) and the Deci-hertz Interferometer Gravitational wave Observatory (DECIGO). For Cosmic Explorer (CE) we use the updated model for the detector noise from \cite{CE}. For the LISA experiment we include the binary confusion noise presented in \cite{LISAConfusion} for our analysis. This stochastic background will also be relevant for the BBO/DECIGO experiments but the precise noise spectrum has not been worked out yet and will depend on the exact detector geometry and the ability to subtract individual sources from the background. Therefore, as a conservative choice we set the lower frequency limit of both detectors to $f=3\times 10^{-3}\,\rm{Hz}$ where we can expect the noise level to be subdominant (see the dotted gray line in Figure~\ref{fig:straintidal}). We found that if we instead extrapolate the detector noise to lower frequencies, including the binary noise as computed for LISA, then our results do not change much. Only if BBO achieves lower detector noise than LISA at frequencies $f\ll10^{-3}$ do the detection prospects improve notably.

    In addition to the noise from the galactic binaries, there are a plethora of other possible sources for GW backgrounds \cite{GWB1,GWB2}, from binary black holes to phase transitions. The existence of a GW background can in principle limit the detectability of the perturbations induced by dark objects flying through the solar system. However, there are often large uncertainties in the estimates for the noise level of the GW backgrounds. Therefore, in our analysis we will assume that detectability is limited only by the detector noise. The only exception to this is the mentioned galactic foreground that is accounted for in the LISA noise curve.

    The integrand in Equation~\ref{eq:snr} has the form $\mathrm{d}f/P_n(f)/f^4=\mathrm{d}\log f/P_n(f)/f^3$ for $f\ll v/R$ and decreases sharply for $f\gg v/R$. It is therefore convenient to plot $\sqrt{P_n(f)f^3}$ and $h(f)f^2$ together, as the latter will be constant for $f\ll v/R$ and the area of the ratio between both curves will be proportional to the SNR on a log-scaled figure.
    
    Figure~\ref{fig:straintidal} shows the rescaled spectral noise of current, planned and proposed GW observatories. This plot can be used to easily read off the signal-to-noise ratio for any of the experiments displayed by the area between the signal, based on the right-side y-axis and the noise curve. The axis have been rescaled such that the signal will be represented by a horizontal line with an amplitude of $\frac{GM}{\pi^2vR(\alpha R+\beta L)}$ that extends up to a frequency $f=v/R$ and drops sharply thereafter. Note that in the the limit $R\ll L$ the amplitude is dependent on the arm length of each detector and thus the signal must be computed for each experiment individually.

    As a first step, we remain agnostic about the orbital parameters of the perturber and compute for a range of values of $M$ and velocity $v$ the distance of the encounter $R$ to which each experiment is sensitive. We will present the results of this in the following section.
    
    As a second step, we study the feasibility of detecting objects with a given interstellar mass density $\rho$ at the location of the solar system. If primordial black holes of a given mass $M$ make up all of the dark matter, they will have $\rho\sim\rho_{\rm DM}=7\times10^{-25}\rm{g\,cm}^{-3}\approx 0.4\,\rm{GeV\,cm}^{-3}$ and an rms velocity of $v\sim 300\,\rm{km\,s}^{-1}$ \cite{DMReview,Choi}. While our results apply to PBHs, they can also be used more generally as a way to constrain any kind of interstellar objects flying through the solar system.
    
    Fixing the density of the perturbers implies a number of encounters with impact parameters smaller than $R$ within a time $t$, given by
    \begin{equation}
        N=\frac{\pi R^2\rho v t}{M}\,.
    \end{equation}
    Therefore, the probability distribution of the distance $R$ to the closest encounter by Poisson statistics is
    \begin{equation}
        p(R)=p(N)\frac{\mathrm{d}N}{\mathrm{d}R}=\frac{2\pi R\rho v t}{M}\exp\left(-\frac{\pi R^2\rho v t}{M}\right)\,,
    \end{equation}
	which implies that the mean distance of the closest encounter to occur within a timespan $t$ is given by
    \begin{equation}
        \langle R_{\rm min}\rangle=\sqrt{\frac{M}{4 \rho vt}}\,.
        \label{eq:meandist}
    \end{equation}
    Using this as an estimate for the value of $R$, we employ Equations~\ref{eq:meanstrain} and \ref{eq:snr} to evaluate the SNR for a given interstellar density $\rho$, mass $M$ and velocity $v$ of the perturbers.

    Previous studies of 'burst'-like GW signals have found SNRs of around 8 sufficient for the LIGO observatory \cite{SearchHyperbolic,FARNoise}. For a lower SNR, the false alarm rate (FAR) will be too large. Using Equation~30 from \cite{FARNoise}, we compute the FAR for any other GW detector using the spectral density $h(\omega)$ of the signal:
    \begin{equation}
        \rm{FAR}=\sqrt{\frac{C_2-C_1^2}{2\pi}}\xi e^{-\xi^2/2}\,.
    \end{equation}
    Here, $\xi$ denotes the signal-to-noise ratio according to Equation~\ref{eq:snr} and the $C_k$ are defined as
    \begin{equation}
        C_k=\frac{4}{\xi^2}\int\mathrm{d}f\,(2\pi f )^k\frac{h(f)^2}{P_n(f)}\,.
    \end{equation}
    As the detectors considered in this work are generally most sensitive at lower frequencies compared to LIGO, the resulting false alarm rates are generally lower and so is the required SNR.
        
	\section{\label{sec:results}Detection prospects} 

    \begin{figure*}
        \centering
        \includegraphics[width=\linewidth]{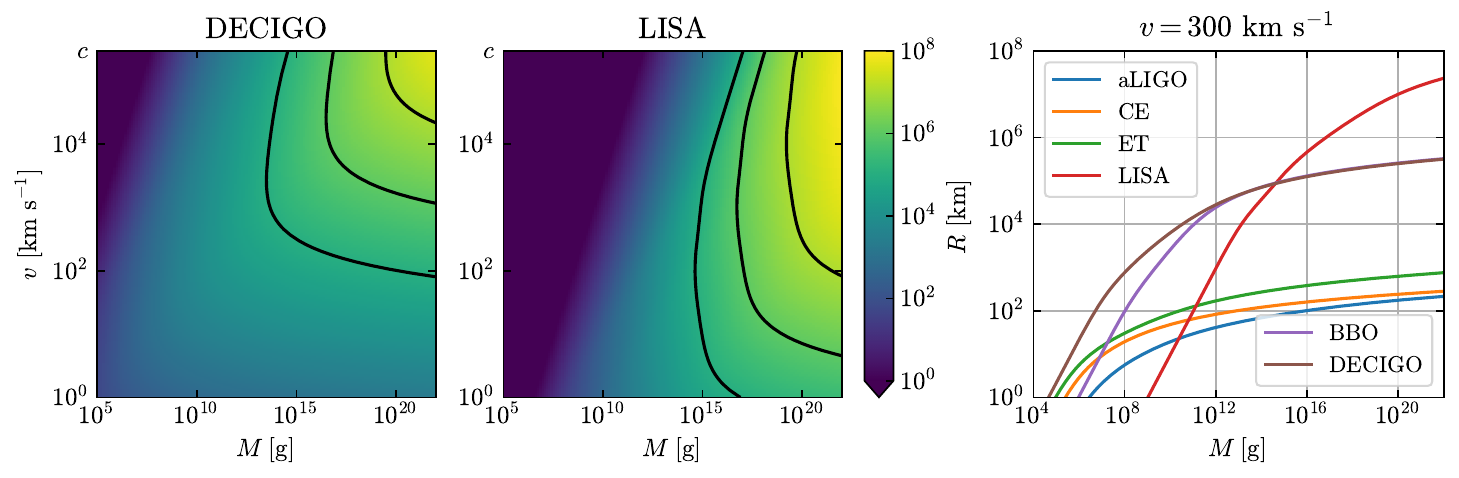}
        \caption{Maximum distance for which GW experiments are sensitive to a fly-by of a perturber. Left: Results for DECIGO/LISA as a function of the mass and velocity of the perturber. Contours for $R=[10^5,10^6,10^7]\,\rm{km}$ are shown as black lines. Right: Results for all GW experiments as a function of the mass of the perturber for $v=300\,\rm{km\,s}^{-1}$. Detectability is defined as $\rm{FAR}<10^{-3}\,\rm{yr}^{-1}$.}
        \label{fig:dist}
    \end{figure*}

    Figure~\ref{fig:dist} displays the minimum fly-by distance required to have a signal with $\rm{FAR}<10^{-3}\,\rm{yr}^{-1}$, which corresponds to SNRs between 5 and 8 for the experiments considered. We display the detection volume as a function of the perturber mass and velocity for LISA and DECIGO and additionally show the results for all detectors for $v=300\,\rm{km\,s}^{-1}$.
    
    A limitation for each GW detector is the minimum frequency $f_{\rm min}$ that it is sensitive to, as it sets the maximum distance observable to $R_{\rm max}\sim v/f_{\rm min}$. For this reason, LISA has the largest detection range for the most massive objects due to its susceptibility at low frequencies. For objects of smaller mass or higher velocity, where the limiting factor is the sensitivity of the experiment, DECIGO achieves a larger detection volume due to its superior design sensitivity compared to LISA.

    We note that the large detection radius for LISA in principle opens the possibility to measure the masses of asteroids or comets whose trajectory and velocity is known, should they pass close enough to the detector. The recently discovered interstellar object 3I/ATLAS \cite{3Atlas,3AtlasLoeb,3AtlasJPL} has a minimum intersection distance with Earth's orbit of 0.37~AU. Assuming a mass of $10^{16}\,\rm{g}$, corresponding to a diameter of the nucleus of 1-2 km, the closest approach to LISA would have to be at a distance below 0.002~AU in order to be detectable. However, there are assumed to be many more smaller bodies, some of which could come close enough to a GW observatory for a detection. Computing the likelihood for such an event is left for future work.

    \begin{figure*}
        \centering
        \includegraphics[width=\linewidth]{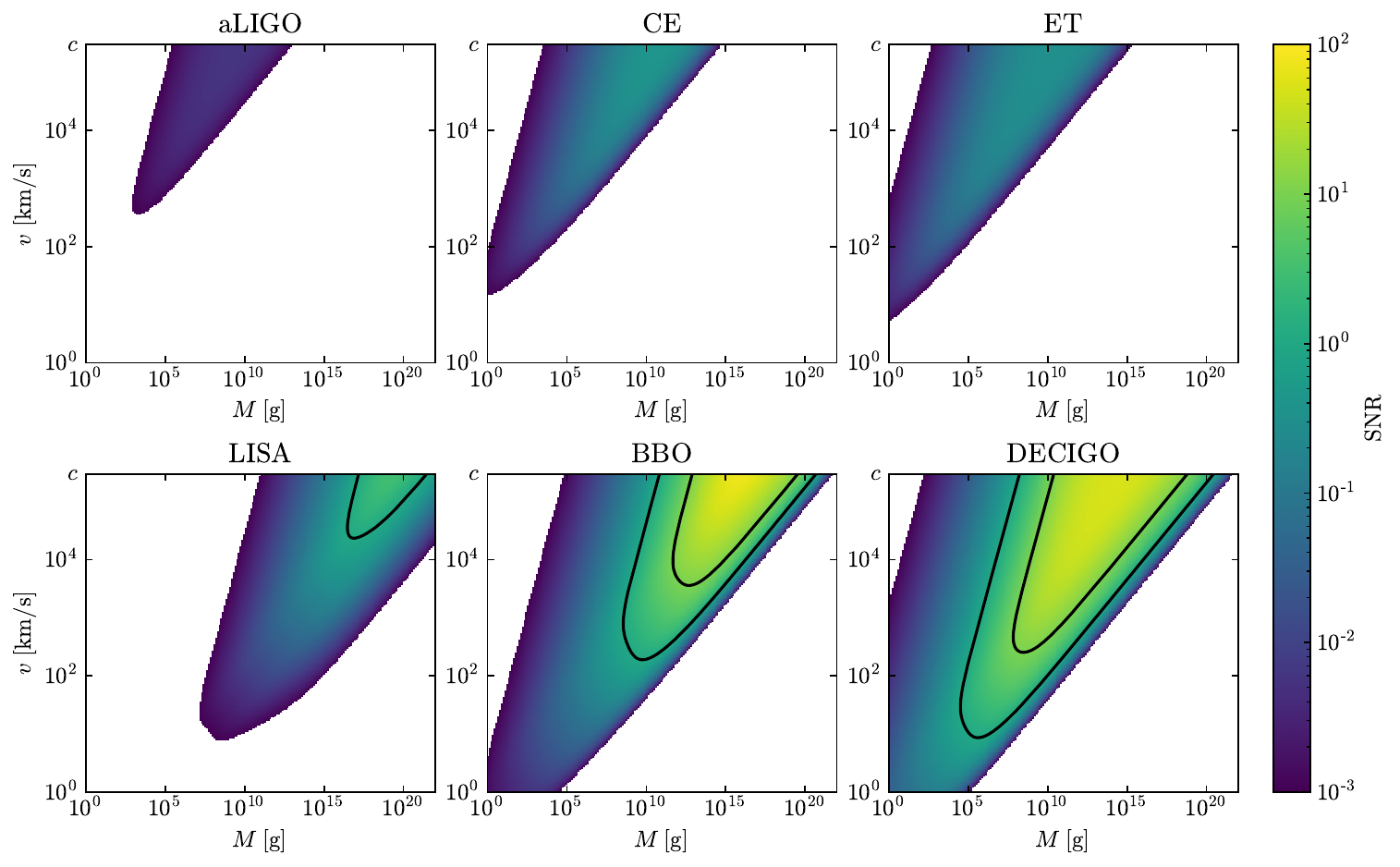}
        \caption{Signal-to-noise ratio from the closest encounter expected within an observation time of ten years and for a density $\rho=\rho_{\rm DM}=7\times10^{-25}\,\rm{g\,cm}^{-3}$. The SNR is shown for several GW experiments as a function of the mass $M$ and velocity $v$ of the perturber. In the white area the SNR is lower than $10^{-3}$. The black lines indicate the contours for which $\rm{SNR}=[1,10]$.}
        \label{fig:snr}
    \end{figure*}

    Figure~\ref{fig:snr} shows the SNR resulting from a fixed density of the dark objects of $\rho\sim\rho_{\rm DM}=7\times10^{-25}\rm{g\,cm}^{-3}\approx 0.4\,\rm{GeV\,cm}^{-3}$. The distance to the perturber is set by Equation~\ref{eq:meandist}. The future detectors BBO and DECIGO show the most promising results, achieving large SNRs for a sizable part of the parameter space, even at non-relativistic velocities. LISA, CE and ET reach SNRs close to or slightly above one at relativistic velocities. However, such parameter regions are likely ruled out by other constraints, either from structure formation in the case of dark matter or from collisions with solar system bodies in general.

\begin{figure}
        \centering
        \includegraphics[width=\linewidth]{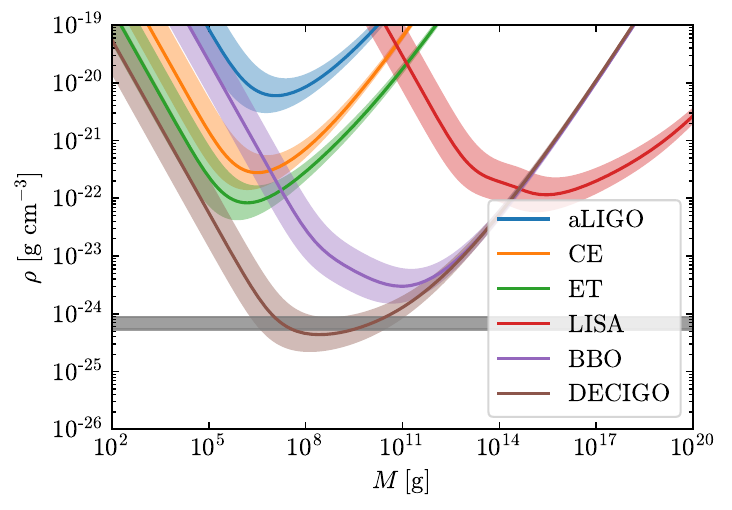}
        \caption{Density to which each experiment is sensitive at a velocity of $v=300\,\rm{km\,s}^{-1}$ as expected for dark matter. The results are obtained by requiring $\rm{FAR}<10^{-3}\,\rm{yr}^{-1}$ which implies SNRs between 5 and 8 for the various detectors. The colored shaded area indicates the resulting range implied by an uncertainty in the detector noise $\sqrt{P_n}$ of a factor of two. The gray shaded area corresponds to the range of the latest observational estimates of the local dark matter density $\rho_{\rm DM}\in[0.3,0.5]\,\rm{GeV\,cm}^{-3}$ \cite{DMReview}.}
        \label{fig:rhodm}
    \end{figure}

    Focusing on the case of compact objects such as PBHs as dark matter, we adopt a velocity of $v=300\,\rm{km\,s}^{-1}$ and compute the lowest density $\rho$ for which the closest encounter within a period of $T=10\,\rm{yr}$ produces a signal with a false alarm rate below $10^{-3}\,\rm{yr}^{-1}$. The results are displayed in Figure~\ref{fig:rhodm}. The curves can be easily adopted for different periods of observation $T$ by noting that $\rho\sim 1/T$. Our results demonstrate that the planned BBO and DECIGO experiments will be able to probe scenarios where the dark matter is in the form of compact objects with a mass of $M\in[10^7,10^{11}]\,\rm{g}$. It is commonly assumed that PBHs of such masses cannot make up the dark matter due to their rapid evaporation. However, the so-called 'memory burden' effect \cite{Dvali2018,Dvali2020} could significantly suppress the evaporation, opening a new window for light PBHs as a dark matter candidate \cite{Alexandre,Thoss,Montefalcone,Zell}. It is therefore highly interesting that GW experiments could be used to probe their existence purely by their gravitational interaction as they pass through the solar system. By pure coincidence, DECIGO is also able to probe PBHs of similar mass through the GW emission associated with their formation \cite{Barker}.

    The noise spectrum of future gravitational wave detectors has to be estimated, which implies some degree of uncertainty. In Figure~\ref{fig:rhodm} we show how an uncertainty in the detector noise $\sqrt{P_n(f)}$ of a factor of two affects the resulting densities which can be probed. For lower masses, where the detection is limited by the noise level of the observatory one has $\rho \sim P_n(f)$ as $h(f)\sim 1/R\sim \sqrt{\rho}$. At higher masses, where the limit is set by the frequency range of the detector, the results are not very sensitive to the exact noise level.

    Our results show that DECIGO has the greatest potential to detect perturbations from dark matter in the form of compact objects streaming through the solar system. We therefore study this case in more detail and relax previous approximations. For a more accurate assessment of detection prospects we compute the signal using Equation~\ref{eq:acc_general} by performing Monte Carlo simulations over the space of possible geometries of the encounter. The velocity for each encounter is obtained by adding a random component with $v_{\rm rms}=270\,\rm{km\,s}^{-1}$ to the proper motion of of the sun around the galactic center (DM wind) with $v_\odot=220\,\rm{km\,s}^{-1}$ \cite{Choi}. We neglect the small non-isotropy in the angular distribution of the velocity and sample the orientation uniformly on a sphere. These simulations are performed for a range of different masses $M$ and densities $\rho$ and the probability of a successful observation in a period of $T=10\,\rm{yr}$ is determined as the fraction of simulations for which $\rm{FAR}<10^{-3}\,\rm{yr}^{-1}$ is obtained. 

    \begin{figure}
        \centering
        \includegraphics[width=\linewidth]{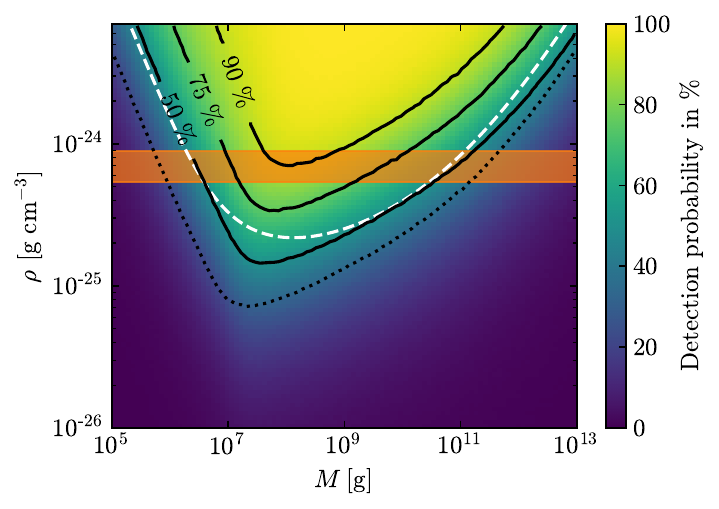}
        \caption{Probability for DECIGO to detect at least one signal from the fly-by of compact objects, shown as a function of their mass $M$ and density $\rho$. We assume an observation time of $T=10\,\rm{yr}$. A successful detection is defined as having $\rm{FAR}<10^{-3}\,\rm{yr}^{-1}$, which roughly corresponds to $\rm{SNR}>6$. The orange shaded area corresponds to the range of the latest observational estimates of the local dark matter density $\rho_{\rm DM}\in[0.3,0.5]\,\rm{GeV\,cm}^{-3}$ \cite{DMReview}. Contours of constant probability are displayed as solid black lines. The dotted black line indicates a contour of 50\% probability given a reduction in the detector noise level by a factor of two. The estimate from Figure~\ref{fig:rhodm} for the DECIGO experiment is shown as a dashed white line.}
        \label{fig:probdecigo}
    \end{figure}
    
    Figure~\ref{fig:probdecigo} shows the results. Notably, there is a wide range of masses $M$ for which the probability of observing the gravitational signal is high. In other words, should the dark matter be in the form of compact objects with mass $M\in[10^7,10^{11}]\,\rm{g}$, there is a good chance that it will perturb the DECIGO detector enough to lead to a detectable signal at least once within 10 years of observation. This agrees well with the estimate presented before in Figure~\ref{fig:rhodm} that is also shown as a white dashed line in Figure~\ref{fig:probdecigo}. The non-detection of a gravitational signal of this kind allows in principle to put constraints on the dark matter density for the same range of masses. However, such bounds might not have a very high confidence, less than 2 sigma based on our results. 
    
    The sensitivity to the perturbations depends on the exact level of DECIGO's detector noise. We highlight this in Figure~\ref{fig:probdecigo} by an additional dashed line which indicates the contour for a 50\% probability of observation in a scenario where the detector noise is lower by a factor of two. Notably, if the 'Ultimate DECIGO' detector \cite{uDECIGO} is realized, which is only limited by quantum noise, then the detection prospects could increase by several orders of magnitude. 
    Conversely, the detection prospects can diminish if detector noise is greater than anticipated or in the presence of irremovable gravitational wave backgrounds. We find that a factor of 5 increase in the noise level reduces the probability for an observable to below 50\% for any mass of the perturber. Finally, we want to note that the densities which are observable or that can be constrained scale with observation time as $\rho\sim 1/T$.
    
	\section{\label{sec:summary}Summary} 

    We have studied the feasibility of gravitationally probing the streaming of dark objects through our solar system. The detection method relies on the perturbation of the test masses of gravitational wave detectors, which leads to a burst-like signal in the experiment.

    We have computed the distances of encounters to which the different types of current and future GW experiments are sensitive, which can reach several million kilometers for LISA, BBO, and DECIGO. The main limitation of each experiment is the frequency range to which it is sensitive. In order for the encounter to be detectable, the fly-by must be sufficiently close or fast in order for $f\sim v/R$ to be large enough.

    If the solar system is penetrated by dark objects with a halo density $\rho\sim\rho_{\rm DM}$ then future GW experiments such as BBO and DECIGO offer the best prospects to detect dark matter in a mass window  $M\in[10^7,10^{11}]\,\rm{g}$. Such a scenario is highly interesting for the case of PBHs, whose evaporation is suppressed by the 'memory burden' effect.
    
	\section{acknowledgments}
		We want to thank Valeriya Korol, Jakob Stegmann and Andreas Burkert for helpful comments. This research was supported in part by the Excellence Cluster ORIGINS which is funded by the Deutsche Forschungsgemeinschaft (DFG, German Research Foundation) under Germany’s Excellence Strategy - EXC-2094 - 390783311 (for V.T.). This work was supported in part by the Black Hole Initiative and the Galileo Project at Harvard University (for A.L.).

    \section{Data availability}
        The numerical results presented in Figure~\ref{fig:dist} and Figure~\ref{fig:rhodm} are publicly accessible under \url{https://doi.org/10.5281/zenodo.16411495} \cite{Paperdata}. Any other data is available upon request.

\bibliography{apssamp}

\end{document}